\documentclass[a4paper, 11pt]{article}

\usepackage[utf8]{inputenc} 
\usepackage[T1]{fontenc}    
\usepackage[english]{babel} 
\usepackage{amsmath}        
\usepackage{graphicx}       
\usepackage{authblk}        
\usepackage[margin=1in]{geometry} 

\usepackage{subcaption} 
\usepackage[colorlinks=true, citecolor=magenta, linkcolor=blue, urlcolor=blue]{hyperref}
\usepackage{siunitx}    
\usepackage{graphicx}
\usepackage{hyperref}
\usepackage{cleveref}
\hypersetup{
    colorlinks=true,
    linkcolor=blue,
    filecolor=magenta,      
    urlcolor=cyan,
    pdftitle={Your Paper Title},
    pdfpagemode=FullScreen,
}

\usepackage[utf8]{inputenc}
\usepackage[T1]{fontenc}
\usepackage{xcolor} 
\usepackage{authblk}

\usepackage[normalem]{ulem}

\title{\textbf{3D-Deuteron Track Recoils Produced by Neutron Capture in Hydrogen Measured by MIMAC-35 cm}}

\author[1]{I. Ourahou}
\author[1]{D. Santos}
\author[1]{O. Guillaudin}
\author[1]{P. Louis-Cistac}
\author[1]{F. Malek} 
\author[1]{N.~Sauzet}
\author[2]{C. Tao}

\affil[1]{LPSC, Université Grenoble Alpes, CNRS/IN2P3, Grenoble, France}

\affil[2]{Centre de Physique des Particules de Marseille, IN2P3/CNRS and Université Aix-Marseille,
Marseille, France}
\affil[ ]{E-mail: \textcolor{violet}{ilias.ourahou@lpsc.in2p3.fr}, \textcolor{violet}{daniel.santos@lpsc.in2p3.fr}}
\date{}

\begin{document}

\maketitle

\begin{abstract}
The neutron capture is a process that concerns most of the nuclei used to build our detectors. This process produces protons, alpha particles, and gamma rays which generate background signals. Characterizing this background is important for rare event searches, such as dark matter detection or Coherent Elastic Neutrino-Nucleus Scattering (CEvNS). This paper presents the result of the direct measurement of thermal neutron captures in hydrogen using a new MIcro-TPC MAtrix Chamber (MIMAC-35 cm) detector with a sensitive volume of 35$\times$ 35$\times$29~cm$^3$. Data were collected over more than 5 days (443519 sec) with a gas mixture at 30 mbar of 70\% isobutane (C$_4$H$_{10}$) and 30\% trifluoromethane (CHF$_3$). Our discrimination method is based on using 3D tracks and released ionization energy, in order to discriminate nuclear recoils (NR) from the dominant electron recoil (ER) background. This method enables the clear identification of 1.3 keV deuteron tracks resulting from the nuclear capture reaction $^{1}\text{H}(n, \gamma)^{2}\text{H}$. We observed $51$ neutron capture events among more than 11 million total events mainly produced by muons in the experimental room of our ground laboratory. In parallel we have measured the thermal neutron flux just below the chamber with a BF$_3$ detector and a simulation has been performed to estimate the number of captures expected. This work shows the discrimination power of MIMAC search for low-energy (E $ < $ 1 keV) rare event with a huge background without any shielding.
\end{abstract}
\tableofcontents


\section{Introduction}

The search for rare events, such as the direct detection of Weakly Interacting Massive Particles (WIMPs) for dark matter or the observation of Coherent Elastic Neutrino-Nucleus Scattering (CEvNS), has driven detector technology towards low energy thresholds, often in the sub-keV range. In this low-energy regime, precise understanding and characterization of background sources are essential. Among the most challenging backgrounds are events induced by thermal and fast neutrons. 

Neutrons, originating from cosmic-ray interactions in the atmosphere or natural radioactivity in the surrounding materials, can be thermalized within the experimental setup and subsequently be captured by nuclei in the detector's active volume. In detectors having materials with hydrogen, a dominant capture process is $^{1}\text{H}(n, \gamma)^{2}\text{H}$. This reaction has a Q-value of 2.223~MeV, which is almost entirely carried away by the gamma-ray. This capture reaction produces a deuteron recoil with a well-defined kinetic energy of approximately 1.3 keV.

This recoiling deuteron releases its kinetic energy over a very short path, roughly 750 $\mu m$, creating a dense, compact ionization track characteristic of a nuclear recoil (NR). This signature is particularly problematic as it can perfectly mimic the expected signal from a low-mass WIMP or a CEvNS interaction, thus constituting an irreducible background if not explicitly discriminated. However, the direct measurement of such a low-energy NR is a significant experimental challenge. The 1.3~keV energy produces only a small number of primary electron-ion pairs, requiring a detector with both a very low energy threshold and excellent 3D-angular resolution to resolve the track.

The MIcro-TPC MAtrix Chamber (MIMAC) is a gaseous detector specifically designed to face these challenges. Its ability to perform full 3D reconstruction of particle tracks at a low pressure of $30$ mbar provides a powerful tool for analyzing the 3D-track of low-energy events. This capability allows for an efficient discrimination between dense and short tracks of nuclear recoils and the longer tracks of electron recoils (ER) produced by the gamma rays or muons.

In this paper, we present an analysis dedicated to the identification of the 1.3~keV deuteron signal from thermal neutron captures. We describe the selection criteria, based on both the deposited ionization energy and the 3D track, used to isolate these specific nuclear recoil rare events. This work serves not only as a direct measurement of the thermal neutron capture rate in our setup but also as a crucial demonstration of the MIMAC detector's performance in the challenging energy range around 1~keV.

\section{Experimental setup}

We used the new MIMAC-35cm detector \cite{Guillaudin2025} installed at the LPSC in Grenoble, France. The detector is a gaseous micro Time Projection Chamber (µ-TPC) with a sensitive volume of 35$\times$35$\times$29~cm$^3$. It features a 29~cm drift region, and a signal amplification based on a Micromegas detector with a gap of 512~$\mu$m. The readout anode is segmented into 896 strips on each axis (X and Y), corresponding to a pitch of  390~$\mu$m.  For this measurement, the detector was filled with a gas mixture of 70\% isobutane (C$_4$H$_{10}$) and 30\% trifluoromethane (CHF$_3$) at a low pressure of 30~mbar. These operating conditions are optimized for the detection and 3D reconstruction of low-energy nuclear recoils, such as those expected from low energy (8 keV) neutron spectral measurements \cite{beaufort_2021}.
\begin{figure}[ht!]
   \centering
   
   \includegraphics[width=\textwidth]{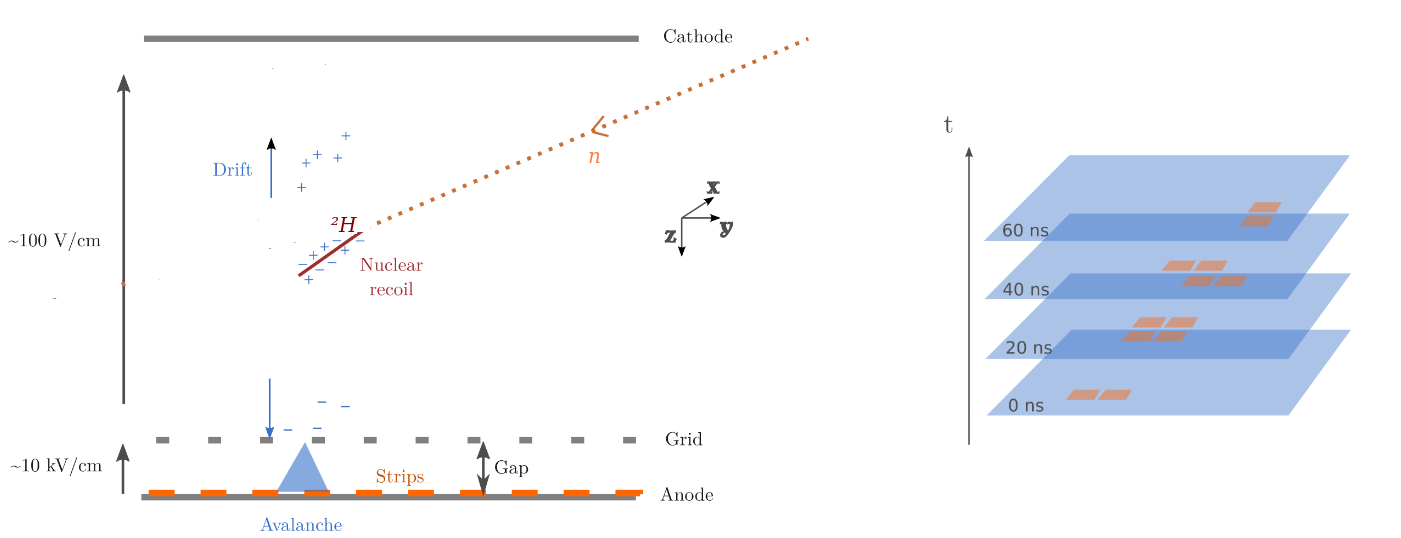}
   \caption{The strategy principle of MIMAC detector.~\cite{beaufort_2021}}
   \label{fig:detector}
\end{figure}

The operating principle of the detector relies on the ionization of the gas by the particle and radiation interacting with the atoms (electrons and nuclei) of the gas mixture in the active volume. A particle traversing the active volume creates a cloud of primary electron-ion pairs. Under the influence of a uniform electric field in the drift region, the primary electrons drift towards the grid of Micromegas. They then enter into the high-field region of the amplification gap, where they produce electron avalanches by Townsend multiplication. The detector presents a dual readout system:
\begin{itemize}
    \item The total charge from the avalanche in the proportional regime is collected on the Micromegas grid. The integral of this induced current provides a fast signal proportional to the total ionization deposited energy of the event, which is digitized with a sampling of 50 MHz in ADC (Analog-to-Digital Converter) units; this is what we call flash-ADC.
    \item A pixelated anode, located at 512 µm from the grid, collects the charge from the avalanche, providing a 2D projection (X-Y) of the particle's track. The third coordinate (Z) is reconstructed from the arrival time of the electrons at the grid, using the known primary electron drift velocity ($v_d$) and the 20~ns sampling rate of the electronics.
\end{itemize}

This velocity ($v_d$) is determined by the gas composition, pressure, and the applied electric field. We simulated these conditions using the \texttt{Magboltz} software package~\cite{magboltz}. For the experimental parameters of a gas mixture of 70\% isobutane (C\textsubscript{4}H\textsubscript{10}) and 30\% trifluoromethane (CHF\textsubscript{3}) at a pressure of 30~mbar and under a drift electric field of 100~V/cm, the calculated electron drift velocity is:
\begin{equation}
    v_d = 24.36 \pm 0.08  \quad \text{µm/ns}.
\end{equation}
This value allows for a direct conversion of the arrival time of the ionization electrons, sampled in 20~ns time slices, into the Z-axis position with a spatial resolution along the drift axis of approximately 487~µm per time slice. Furthermore, the spatial resolution of the reconstructed tracks is fundamentally limited by the diffusion of primary electrons during their transport toward the grid. This stochastic process leads to a spatial spread of the initial ionization cloud, characterized by the transverse ($D_T$) and longitudinal ($D_L$) diffusion coefficients. Based on the same \texttt{Magboltz} simulations, these coefficients for our specific operating conditions are:
\begin{align}
    D_T &= 399 \pm 18 \quad \text{µm}/\sqrt{\text{cm}}, \\
    D_L &= 464 \pm 26\quad \text{µm}/\sqrt{\text{cm}}.
\end{align}
The resulting spatial spread, defined as $\sigma = D\sqrt{Z}$ where $Z$ is the drift distance, represents the Point Spread Function (PSF) of the detector. For an event occurring at the maximum drift distance of 29~cm, the electron cloud reaches a standard deviation of approximately 2.15~mm in the transverse plane and 2.5~mm along the longitudinal axis. These physical parameters are integrated into our likelihood-based reconstruction algorithm to compute the polar and azimuthal angular spectrum in a diffusion-dominated regime.

\begin{figure}[ht!]
   \centering
   
   \includegraphics[width=0.48\textwidth]{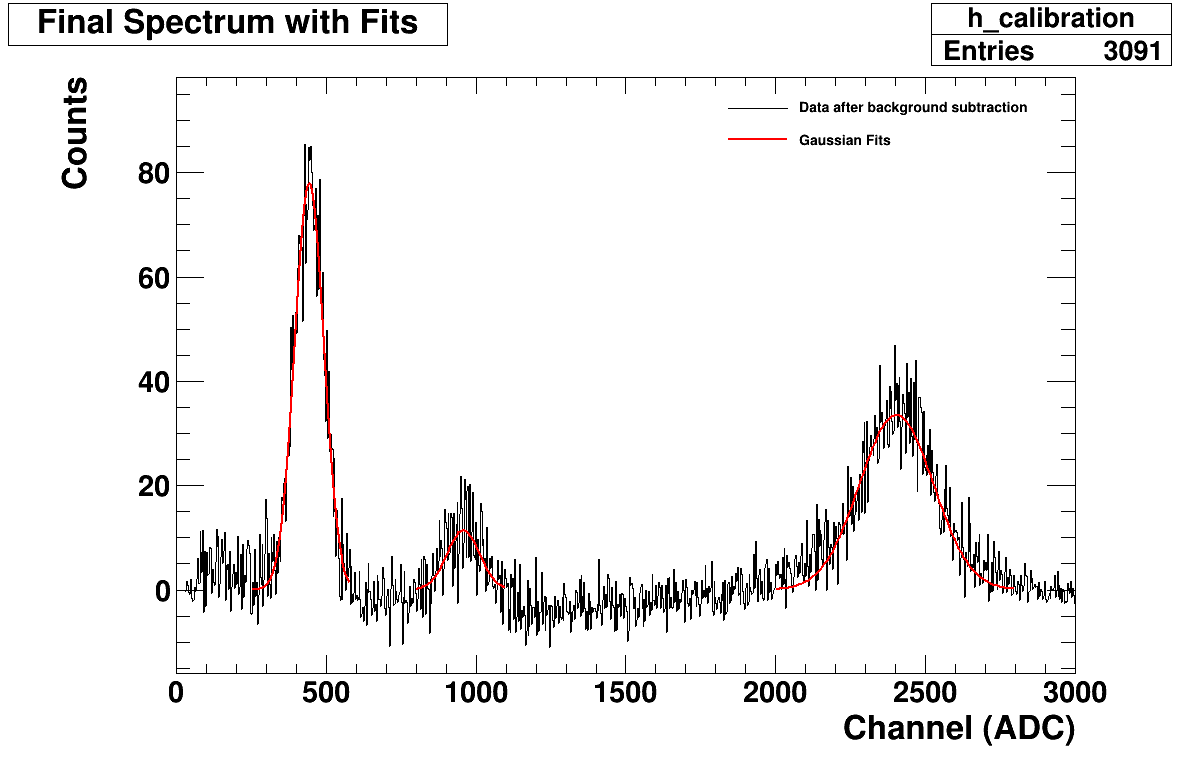}
   \hfill 
   \includegraphics[width=0.48\textwidth]{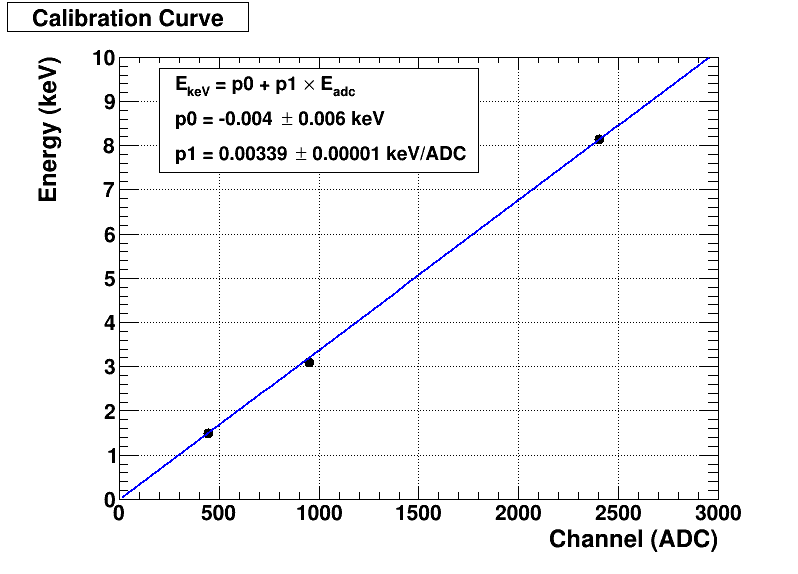}
   
   \caption{Energy calibration of the MIMAC detector. Left: The energy spectrum showing the fluorescence peaks from Al, Cd, and Cu. Right: The resulting linear calibration curve.}
   \label{fig:calibration_complete}
\end{figure}

An energy calibration is performed to convert the measured signal in ADC units to keV electron equivalent. This is achieved using an external X-ray generator that irradiates metallic foils (Aluminum, Cadmium, Copper) placed near the detector. The generator's X-rays induce atomic fluorescence on the metal foils. The resulting energetic fluorescence X-rays then enter the detector's active volume, where they are fully absorbed via the photoelectric effect, creating distinct peaks in the energy spectrum. For this work, we used aluminum (Al K${\alpha}$ and K${\beta}$, effective energy 1.49~keV), cadmium (Cd L${\alpha}$ and L${\beta}$, effective energy 3.16~keV), and copper (Cu K${\alpha}$ and K${\beta}$, effective energy 8.17~keV) foils. A linear fit of the ADC peak positions versus these known energies yields the calibration curve, as shown in Figure~\ref{fig:calibration_complete}. This allows for the determination of the detector's energy threshold, which is found to be below 300~eV. We will see later how the study of these electron events would help us for discriminating them from nuclear recoils.


\subsection{The \texorpdfstring{BF$_3$}{BF3} Detector}

To determine the ambient thermal neutron flux in the experimental area, a measurement was performed using two boron trifluoride (BF$_3$) proportional counters. The two tubes were placed just below the MIMAC detector to ensure that the measured flux was identical to that experienced by the $\mu$-TPC sensitive volume. Each counter is a cylindrical tube, with an active length of 107~cm and a diameter of 5.08~cm, and is housed within a copper casing. The active gas is BF$_3$, enriched to 96\% in the ${}^{10}$B isotope, and the detector was operated at a pressure of 0.93~bar and a temperature of 25$^{\circ}$C. The ${}^{10}$B isotope has a very large thermal neutron capture cross section ($\sigma_{{}^{10}\mathrm{B}} = 3840$~barns at 25~meV). The capture reaction proceeds via two primary channels, producing an $\alpha$ particle and a ${}^{7}\mathrm{Li}$ nucleus whose energies are measured by the detector.

\begin{align}
    n + {}^{10}\text{B} & \rightarrow \alpha + {}^{7}\text{Li}^* \rightarrow \alpha + {}^{7}\text{Li} + \gamma~(0.478~\text{MeV}) \quad (94\% \text{branch, Q = 2.31 MeV}) \\
    n + {}^{10}\text{B} & \rightarrow \alpha + {}^{7}\text{Li} \quad (6\% \text{branch, Q = 2.79 MeV})
\end{align}

\begin{figure}[ht!]
   \centering
   \includegraphics[width=0.48\textwidth]{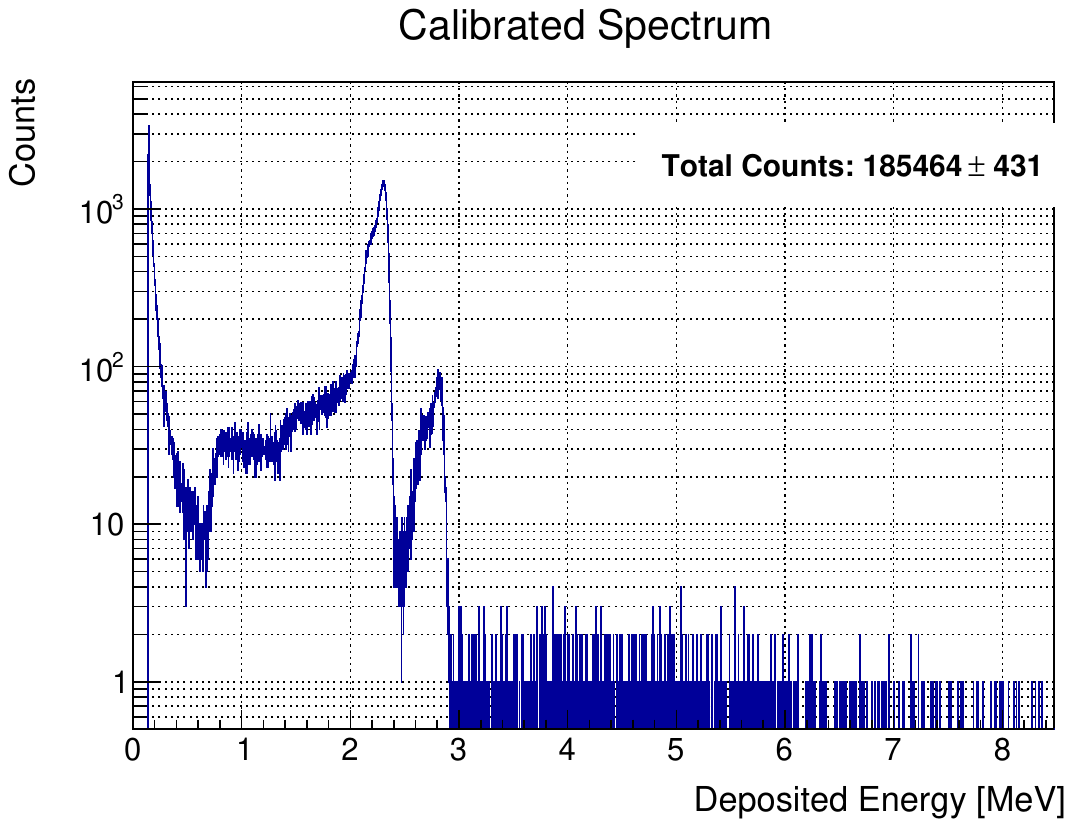}
   \hfill
   \includegraphics[width=0.48\textwidth]{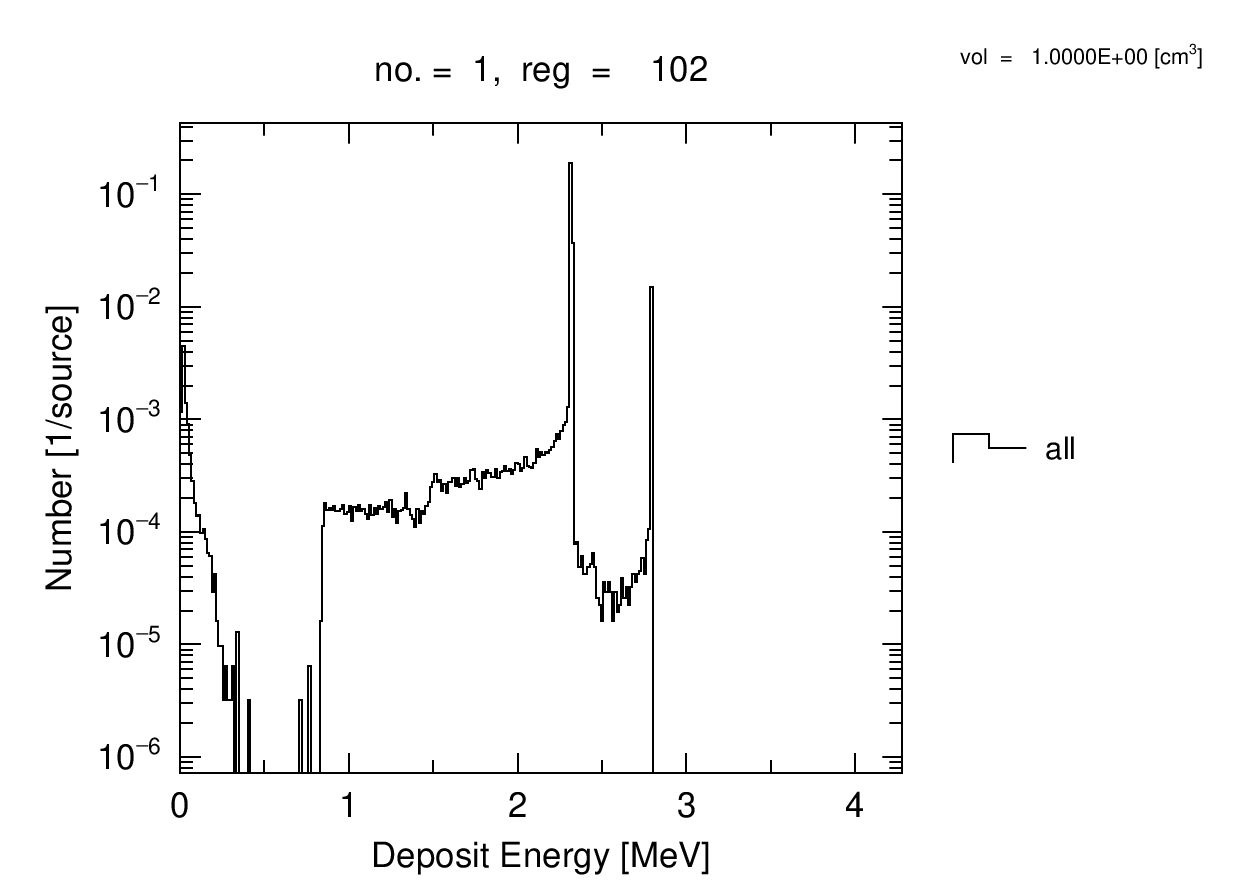}
   \caption{Thermal neutron capture spectrum in the BF$_3$ detector. Left (a): Measured data over 86.15 hours. Right (b): PHITS simulation result, not taking into account the energy detector resolution. The peaks at 2.31 MeV (94\% branch) and 2.79 MeV (6\% branch) are clearly visible in both.}
   \label{fig:bf3_plots}
\end{figure}

Data were collected over a period of $\Delta t = 86.13$~hours. The resulting energy spectrum, shown in Figure~\ref{fig:bf3_plots}(a), exhibits two distinct full-energy peaks corresponding to these two channel reactions. The total number of observed capture events, obtained by integrating the counts within these peaks, is $N_{\text{BF3, exp}} = 144,215$ for both counters, out of a total of 185,464 total events. The identification of these spectral features is confirmed by a PHITS \cite{phits} simulation, shown in Figure~\ref{fig:bf3_plots}(b).

\begin{figure}[ht!]
    \centering
    \includegraphics[width=0.9\textwidth]{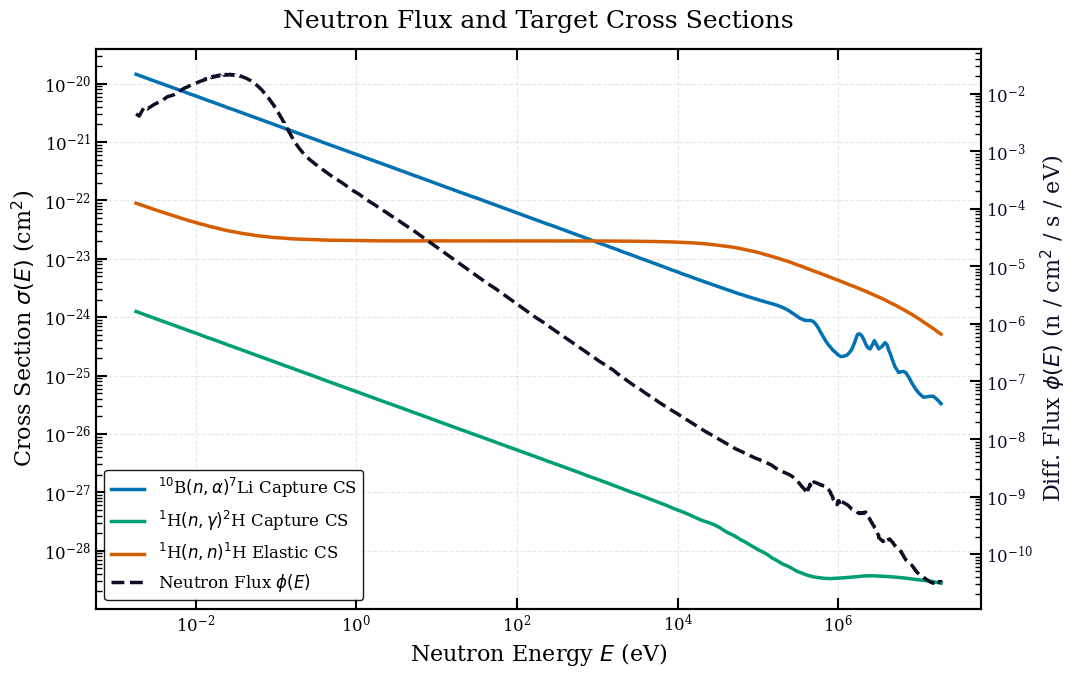} 
    \caption{The local differential neutron flux $\phi(E)$  (red dashed line, right axis) used for the analytical model, shown alongside the capture cross-sections $\sigma(E)$ for $^{10}$B (blue) and $^{1}$H (green) (left axis). The thermal neutron peak overlaps with the region where cross-sections are highest, illustrating that the measured capture rate is dominated by thermalized neutrons. The $^{10}$B cross-section is approximately four orders of magnitude larger than that of $^{1}$H, which accounts for the high statistics obtained in the BF$_3$ reference measurement compared to the MIMAC detector.}
    \label{fig:flux_vs_sigma}
\end{figure}

The primary goal of this measurement is to estimate the local thermal neutron flux. This is achieved by comparing the experimental result with an analytical calculation of the expected number of captures, $N_{\text{calc}}$, given by:
\begin{equation}
N_{\text{calc}} = N_{\text{nucl}} \cdot \Delta t \int_{E_{min}}^{E_{max}} \sigma(E) \cdot \phi_{\text{ref}}(E) \, dE,
\end{equation}

where $N_{\text{nucl}}$ is the total number of target nuclei (${}^{10}$B), $\sigma(E)$ is the energy-dependent neutron capture cross-section~\cite{endf}, and $\phi_{ref}(E)$ is the neutron flux spectrum (Fig.\ref{fig:flux_vs_sigma}). The integral is evaluated numerically over the available energy range of the reference spectrum, from $E_{min} = 1.80 \times 10^{-3}$ eV to $E_{max} = 9.23 \times 10^8$ eV.
For the analytical calculation, we employ a reference spectrum, $\phi_{\text{ref}}(E)$, based on ground-level measurements (at sea level) conducted in New York~\cite{gordon_flux}. To account for the specific geographical and atmospheric conditions of our laboratory in Grenoble, we introduce a scaling factor $F_{\text{alt}}(d)$ proposed by Gordon et al. to correct for the altitude difference. The local neutron flux $\phi_{\text{local}}(E)$ is then expressed as:

\begin{equation}
    \phi_{\text{local}}(E) \approx \epsilon_c \cdot F_{\text{alt}}(d) \cdot \phi_{\text{ref}}(E),
\end{equation}
where $F_{\text{alt}}(d)$ accounts for the predictable enhancement of the atmospheric neutron flux with altitude. It should be noted that the Gordon $\textit{et al.}$ altitude scaling factor was originally derived for high-energy neutrons ($E > 10$ MeV). Furthermore, $\epsilon_c$ is an effective correction factor, describing modifications due to the laboratory environment. In particular, $\epsilon_c$ incorporates neutron reflection and moderation from the walls of the experimental hall, as well as effects related to the local geometry and surrounding materials.

The altitude-dependent enhancement factor is parameterized as
\begin{equation}
    F_{\text{alt}}(d) = \exp \left( \frac{1033.7 - d}{131.3} \right).
\end{equation}
The parameter $d$ represents the atmospheric depth (in g/cm$^2$), calculated from the barometric formula $d = P(h)/g$. The local pressure $P(h)$ at an altitude $h$ is given by:
\begin{equation}
    P(h) = P_0 \cdot (1 - 2.2557 \times 10^{-5} \cdot h)^{5.255}.
\end{equation}
Taking $h = 212$~m as the altitude of Grenoble, we obtain a local atmospheric depth of $d \approx 1008$~g/cm$^2$. Applying this to the scaling function, we find $F_{\text{alt}}(1008) \approx 1.21$, indicating that the neutron flux in Grenoble is approximately 21\% higher than at sea level.

Under this model, the calibration factor $\epsilon_c$ is derived from the ratio of experimental to analytical captures in the BF$_3$ detector, now incorporating the altitude correction:
\begin{equation}
\epsilon_c = \frac{N_{\text{BF3, exp}}}{N_{\text{BF3, calc}}} = \frac{N_{\text{BF3, exp}}}{N_{10\text{B}} \cdot \Delta t \int_{E_{min}}^{E_{max}} \sigma_{10\text{B}}(E) \cdot F_{\text{alt}}(d) \cdot \phi_{\text{ref}}(E) \, dE}.
\end{equation}

Using the calculated number of ${}^{10}$B atoms, the known capture cross-section, and the scaled reference flux, we obtain a calibration factor of $\epsilon_c = 0.7$. Finally, the predicted number of neutron captures on hydrogen in the MIMAC detector is computed by applying this calibrated and scaled flux:
\begin{equation}
N_{c,H}^{\text{pred}} = \epsilon_c \cdot F_{\text{alt}}(d) \cdot N_{1\text{H}} \cdot \Delta t \int_{E_{min}}^{E_{max}} \sigma_{1\text{H}}(E) \cdot \phi_{\text{ref}}(E) \, dE.
\end{equation}

For a data-taking period of 7392 minutes, the revised analytical prediction is:
\begin{equation}
    N_{c,H}^{\text{pred}} = 49 \pm 7 \text{ captures}.
    \label{eq:mimac_prediction_updated}
\end{equation}

\subsection{Monte Carlo Prediction and Comparison with Analytical Model}

To obtain an independent prediction of the neutron capture rate, a Monte Carlo simulation was performed using the PHITS code (version 3.34)~\cite{phits}. The BF$_3$ counter was modeled, and a source of thermal neutrons ($25$~meV) was generated uniformly and isotropically within the volume equivalent to the detector's active region. This volumetric generation mimics the ambient thermal neutron density present in the experimental hall. The total number of simulated projectile neutrons was iteratively adjusted to $N_{\text{BF3}}^{\text{sim}} = 4.8 \times 10^5$ in order to reproduce the $144,215$ captures measured experimentally over the 86.13-hour period. This calibrated value serves as an estimator of the effective thermal neutron population interacting with $^{10} B$  nuclei during the experiment time.

Following this validation, a predictive simulation was performed for the MIMAC TPC. Assuming a homogeneous neutron density in the laboratory, the required number of source neutrons for MIMAC, $N_{\text{H}}^{\text{sim}}$, was determined by scaling the number used in the BF$_3$ simulation by the ratio of the active volumes ($V$) and the exposure times ($\Delta t$):

\begin{equation}
    N_{\text{H}}^{\text{sim}} = N_{\text{BF3}}^{\text{sim}} \cdot \frac{V_{\text{H}}}{V_{\text{BF3}}} \cdot \frac{\Delta t_{\text{H}}}{\Delta t_{\text{BF3}}}.
    \label{eq:neutron_scaling}
\end{equation}

Where $\Delta t_{\text{H}} = 7392$ minutes, this scaling yields an incident source of $N_{\text{H}}^{\text{sim}} \approx 2.8 \times 10^6$ thermal neutrons distributed within the MIMAC volume. Simulating this monoenergetic thermal source (E $\approx$ 25 meV) with the MIMAC detector model predicts a total of:

\begin{equation}
    N_{c,\text{H}}^{\text{sim}} =  61 \pm 8 \text{ captures}.
\end{equation}

\begin{figure}[ht!]
   \centering
   \includegraphics[width=0.6\textwidth]{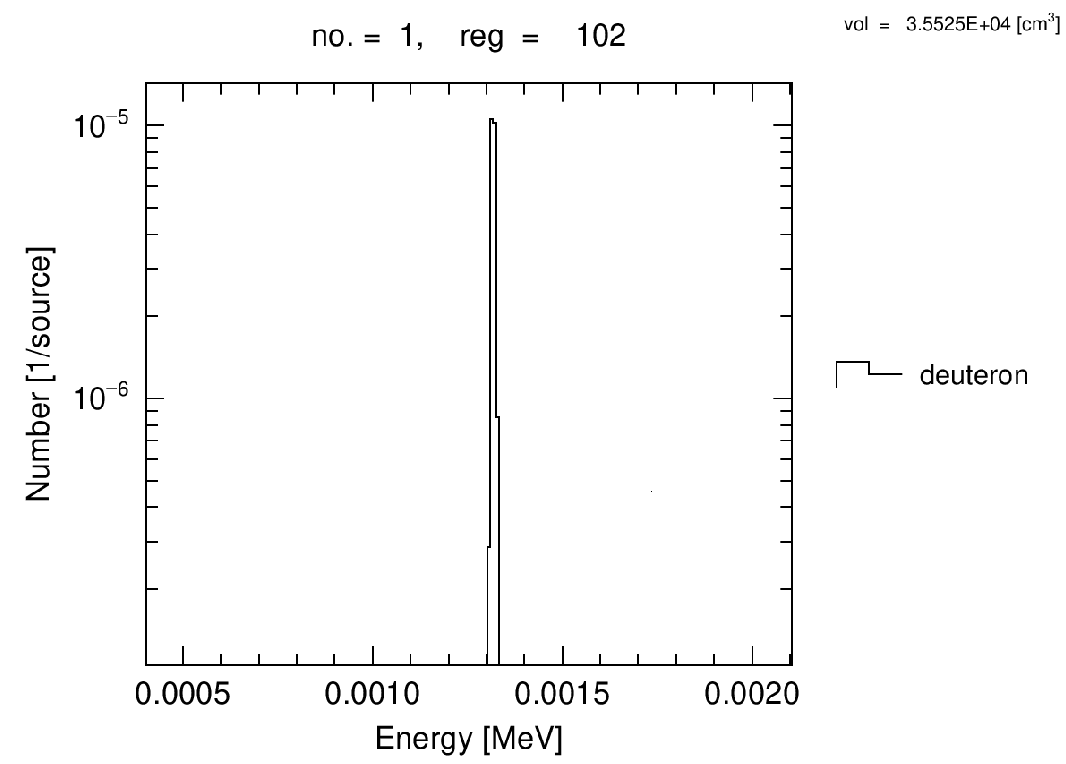}
  
   \caption{PHITS simulation result of the thermal neutron capture spectrum in the MIMAC detector, with a peak at 1.3 keV.}
   \label{fig:h2_plot}
\end{figure}

This simulated capture rate corresponds to events producing a characteristic  deuteron recoil peak at 1.3~keV kinetic energy (Fig.\ref{fig:h2_plot}), it is slightly higher than the analytical prediction ($49 \pm 7$) because we opted to use a strictly monoenergetic 25~meV source in our simulation, which maximizes the capture cross-section. In contrast, the analytical model integrates over the continuous spectrum, extending from the epithermal tail up to the fast neutron regime, where the $1/v$ cross-section significantly decreases. 

\section{Event Selection and Discrimination Strategy}

The discrimination between background nuclear recoils (NR) and the dominant electron recoil (ER) background is achieved through a multi-stage analysis of the ionization signal profile. The selection criteria exploit the distinct physical interaction mechanisms: at equivalent kinetic energies, NRs produce dense, localized ionization tracks due to the higher stopping power of the nuclei than electrons, whereas ERs generate diluted, elongated tracks subject to significant multiple scattering.

\subsection{Data Cleaning and Pre-selection}
\begin{figure}[ht!]
    \centering
    \begin{tabular}{ccc}
    \begin{subfigure}{0.3\textwidth}
        \includegraphics[width=\textwidth]{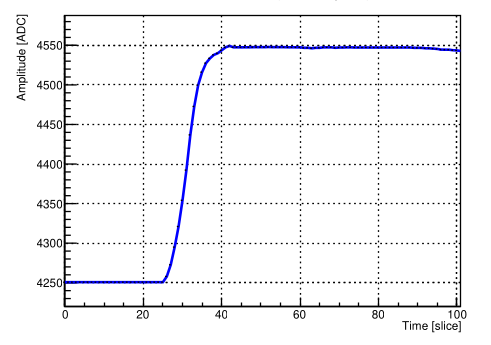}
    \end{subfigure}
    &
    \begin{subfigure}{0.3\textwidth}
        \includegraphics[width=\textwidth]{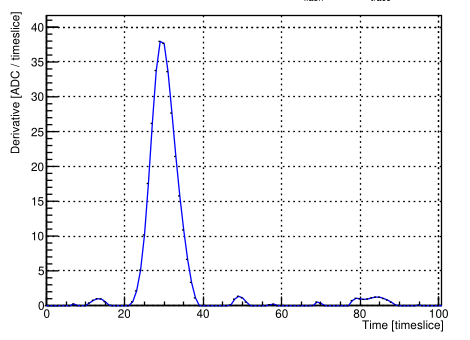}
    \end{subfigure}
    &
    \begin{subfigure}{0.3\textwidth}
        \includegraphics[width=\textwidth]{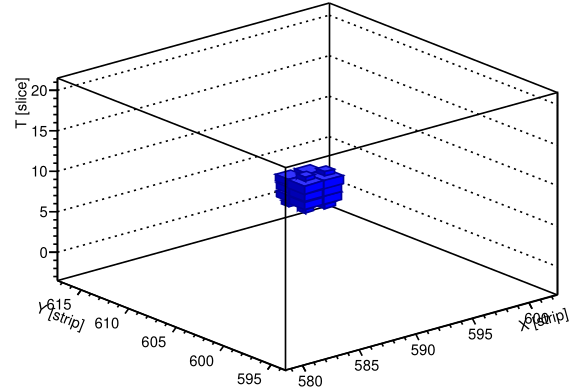}
    \end{subfigure}
    \\
    \begin{subfigure}{0.3\textwidth}
        \includegraphics[width=\textwidth]{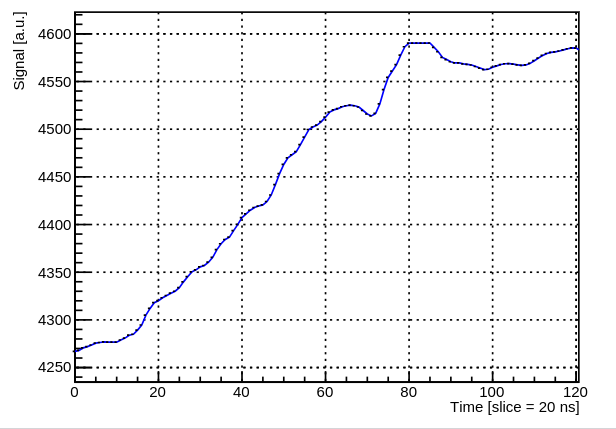}
    \end{subfigure}
    &
    \begin{subfigure}{0.3\textwidth}
        \includegraphics[width=\textwidth]{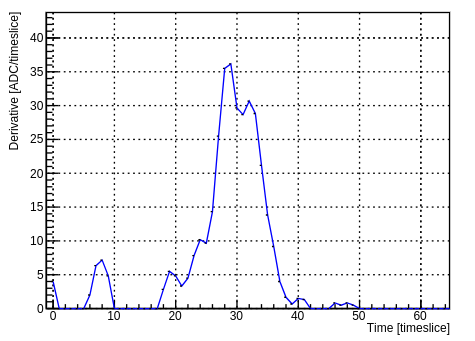}
    \end{subfigure}
    &
    \begin{subfigure}{0.3\textwidth}
        \includegraphics[width=\textwidth]{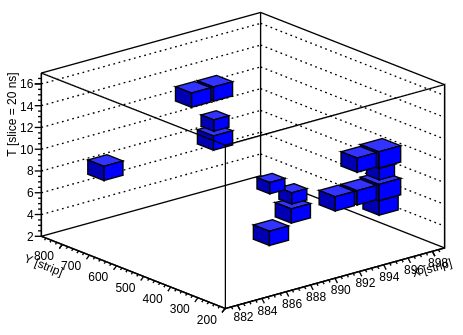}
    \end{subfigure}
    \end{tabular}
    \caption{Comparison of typical event characteristics for a nuclear recoil candidate (top row) and an electron recoil background event (bottom row) at the same energy (E $\approx$ 1~keV). \textbf{Left:} The integrated ionization signal (flash). \textbf{Center:} The temporal derivative of the flash. The nuclear recoil shows a single sharp peak, while the electron recoil exhibits multiple peaks corresponding to distinct ionization clusters. \textbf{Right:} The reconstructed 3D track. The nuclear recoil is a short, point-like track, whereas the electron recoil is diffuse.}
    \label{fig:track_comparison}
\end{figure}

To ensure data quality and isolate physical signals, a pre-selection process is applied based on track topology and signal shape. Two primary filters are applied:
\begin{enumerate}
    \item \textbf{Track Duration:} We require the reconstructed track to span at least two time slices ($2 \times 20$ ns). This criterion effectively suppresses single-point electronic noise and "spark" events that do not exhibit a physical drift time.
    \item \textbf{Flash Derivative Analysis:} The first derivative of the integrated charge signal (Flash-ADC), shown in Fig.\ref{fig:track_comparison}, serves as an observable  to discriminate between nuclear recoils (NR) and electron recoils (ER). We select events exhibiting a single peak in this derivative, corresponding to a single, primary electron cluster. This rejects events with multiple clusters typical of energy (E > 1 keV) electron events.

\end{enumerate}

\subsection*{Topological Analysis: Track Width ($w$)}
To quantify the spatial extension of the primary electron cloud, we define the track width, $w$.  Firstly, we determine the principal axis of each track for each event via a linear regression performed on the 3D coordinates of the cluster. The width observable is calculated as the root mean square (RMS) of the orthogonal distances of all activated pixels relative to the track's principal axis. Mathematically, $w$ is expressed as:

\begin{equation}
    w = \sqrt{\frac{1}{N}\sum_{i=1}^{N} d_i^2},
\end{equation}

\noindent where $N$ is the total number of activated pixels and $d_i$ represents the residual orthogonal distance of the $i$-th pixel to the fitted axis.

\begin{figure}[htbp]
    \centering
    \begin{subfigure}[b]{0.48\textwidth}
        \centering
        \includegraphics[width=0.8\textwidth]{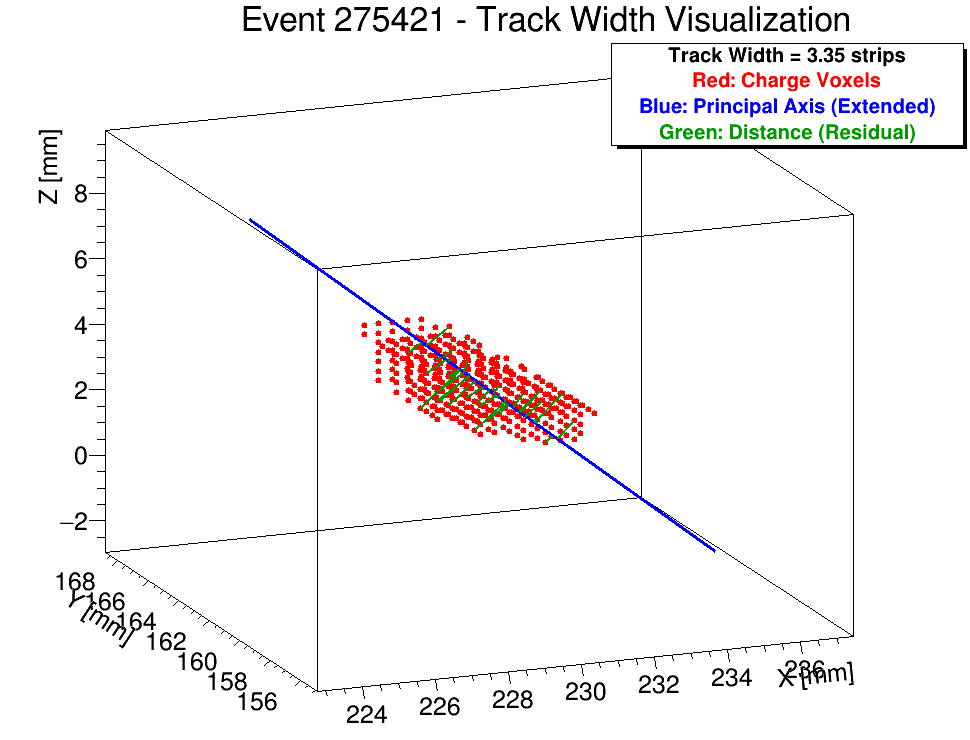}
        \includegraphics[width=0.5\textwidth]{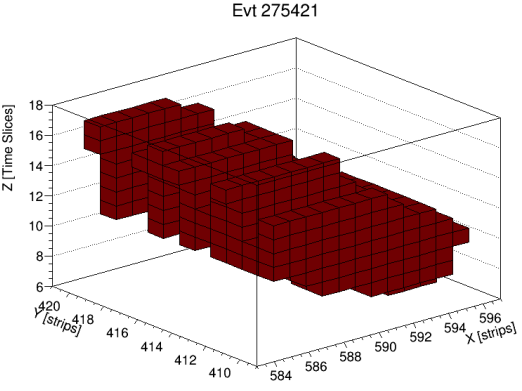}
        \caption{Nuclear Recoil Candidate (4.09 keV)}
    \end{subfigure}
    \hfill
    \begin{subfigure}[b]{0.48\textwidth}
        \centering
        \includegraphics[width=0.8\textwidth]{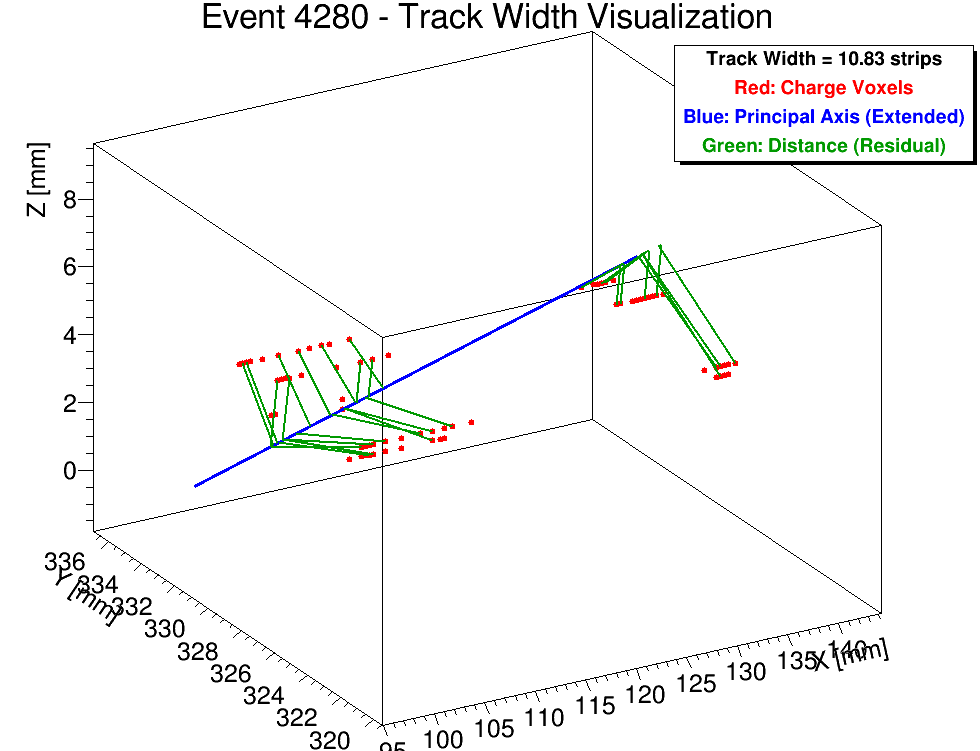}
        \includegraphics[width=0.5\textwidth]{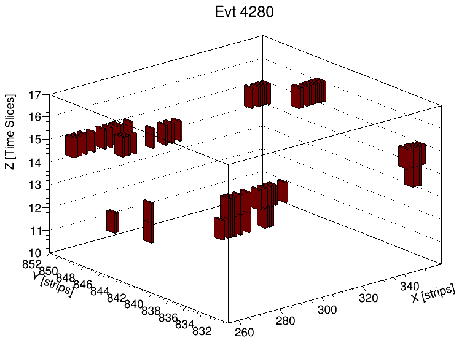}
        \caption{Electron Recoil Background (4.59 keV)}
    \end{subfigure}
    \caption{Topological comparison of reconstructed tracks. The green lines represent the projection of pixel positions onto the principal axis (blue). \textbf{(a)} The nuclear recoil exhibits a compact structure with small lateral deviations. \textbf{(b)} The electron recoil shows a larger spread due to multiple scattering, resulting in a larger width $w$.}
    \label{fig:width_comparison}
\end{figure}

The rejection limits for $w$ are established using a dedicated calibration dataset composed exclusively of electron recoils (X-rays). As illustrated in Figure~\ref{fig:width_comparison}, electrons undergo significant multiple scattering, resulting in systematically larger track widths compared to nuclear recoils of the same energy. Consequently, the distribution of $w$ for electrons defines the conservative upper bound for the acceptance region of nuclear recoils.

\begin{figure}[ht!]
    \centering
    \includegraphics[width=0.8\textwidth]{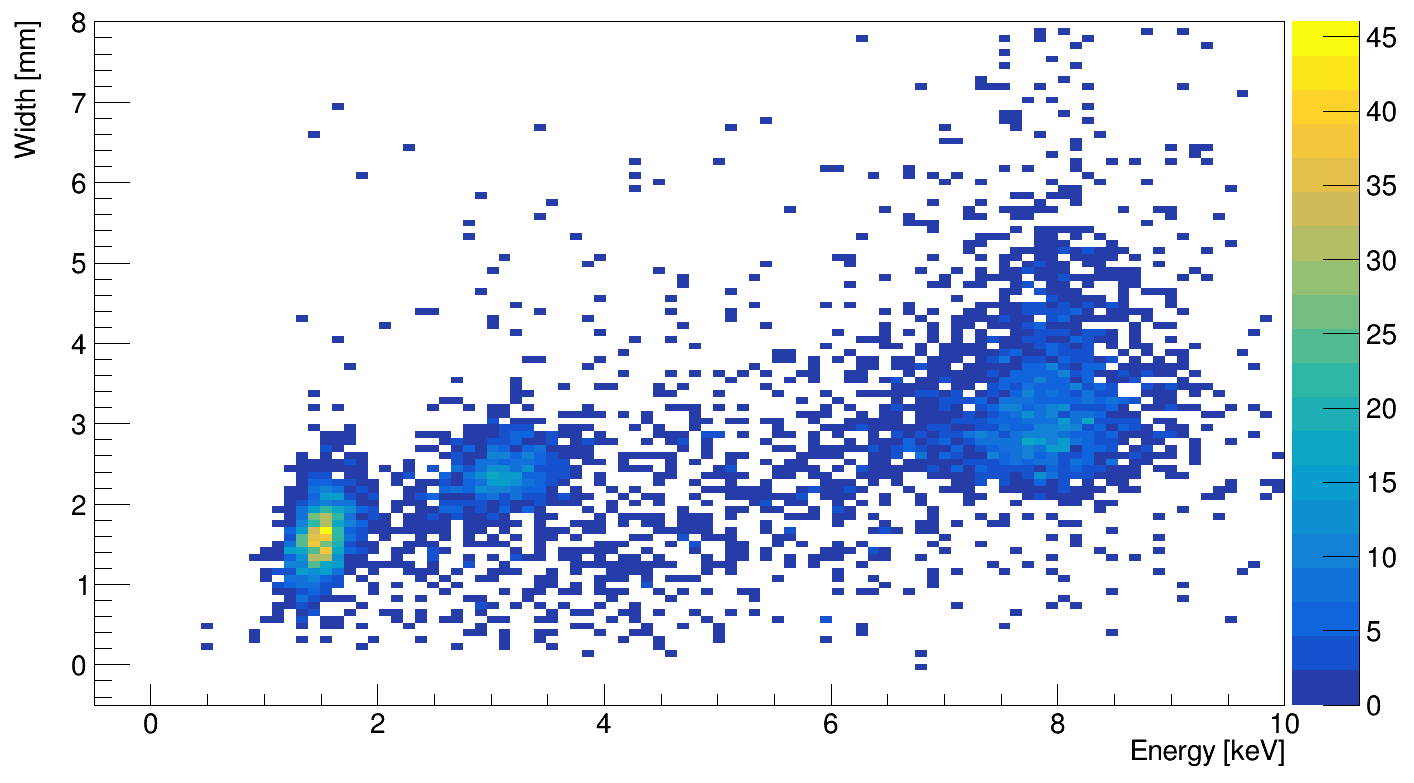} 
    \caption{Distribution of track width $w$ versus energy for the electron recoil calibration dataset. The observed population allows for the definition of an upper exclusion limit for the nuclear recoil search window.}
    \label{fig:bidim_width}
\end{figure}

Figure~\ref{fig:bidim_width} presents the distribution of track width as a function of ionization energy for the electron calibration data. Based on these distributions, we define energy-dependent cuts to maximize the rejection of the ER population while preserving the NR signal acceptance region: $E < 2$ keV and $w < 1$  mm.

We opted for a cut-based analysis rather than a BDT, as the physical cuts provide a clear and sufficient separation between the signal and the background.

\subsection*{Ionization Density Analysis}
Complementary to the track width, the ionization density provides a powerful discriminant. Nuclear recoils create a higher density of primary electron-ion pairs per unit path length. For each event, we introduce the pixel density observable $R$, defined as the number of activated pixels normalized by the  energy ($R \propto N_{pix}/E$).

\begin{figure}[ht!]
    \centering
    \includegraphics[width=0.45\textwidth]{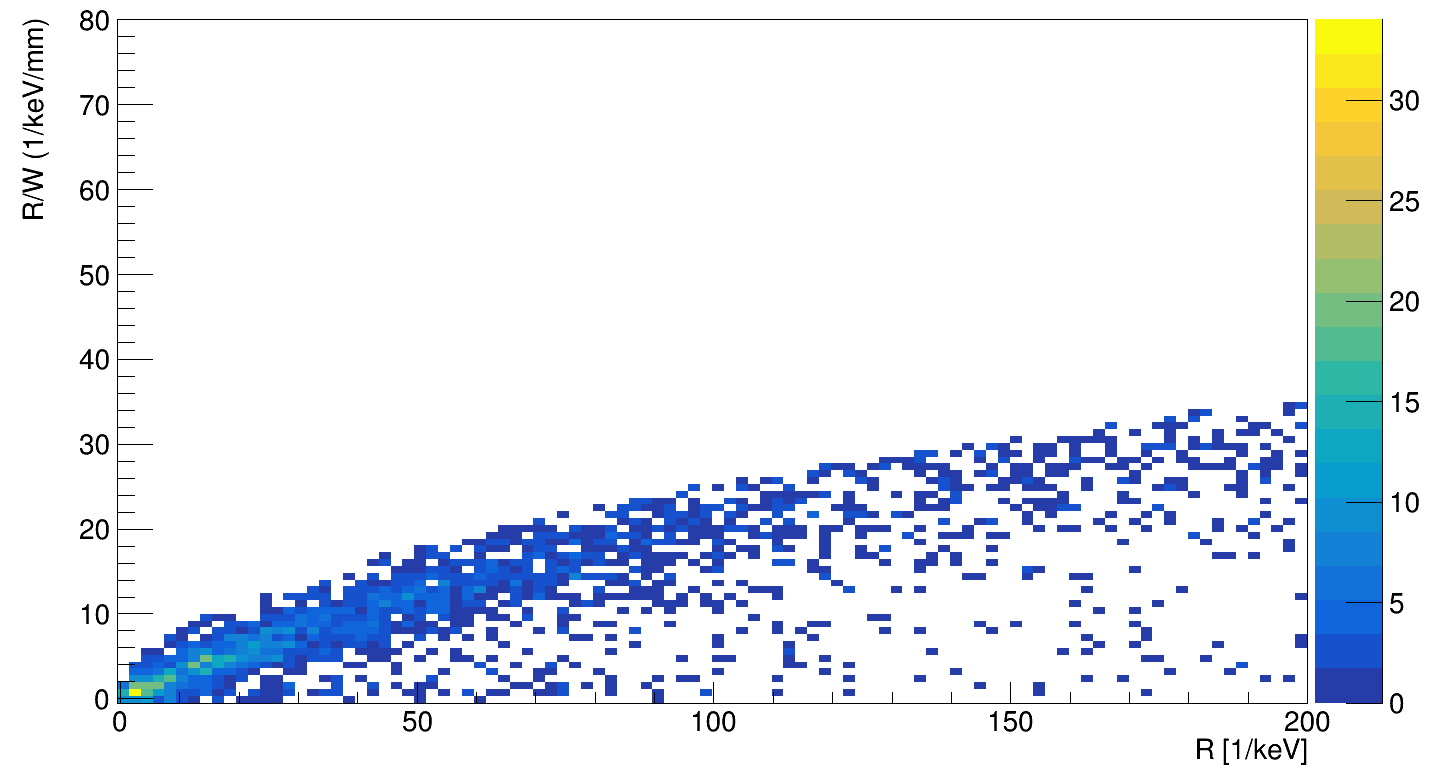} 
    \includegraphics[width=0.45\textwidth]{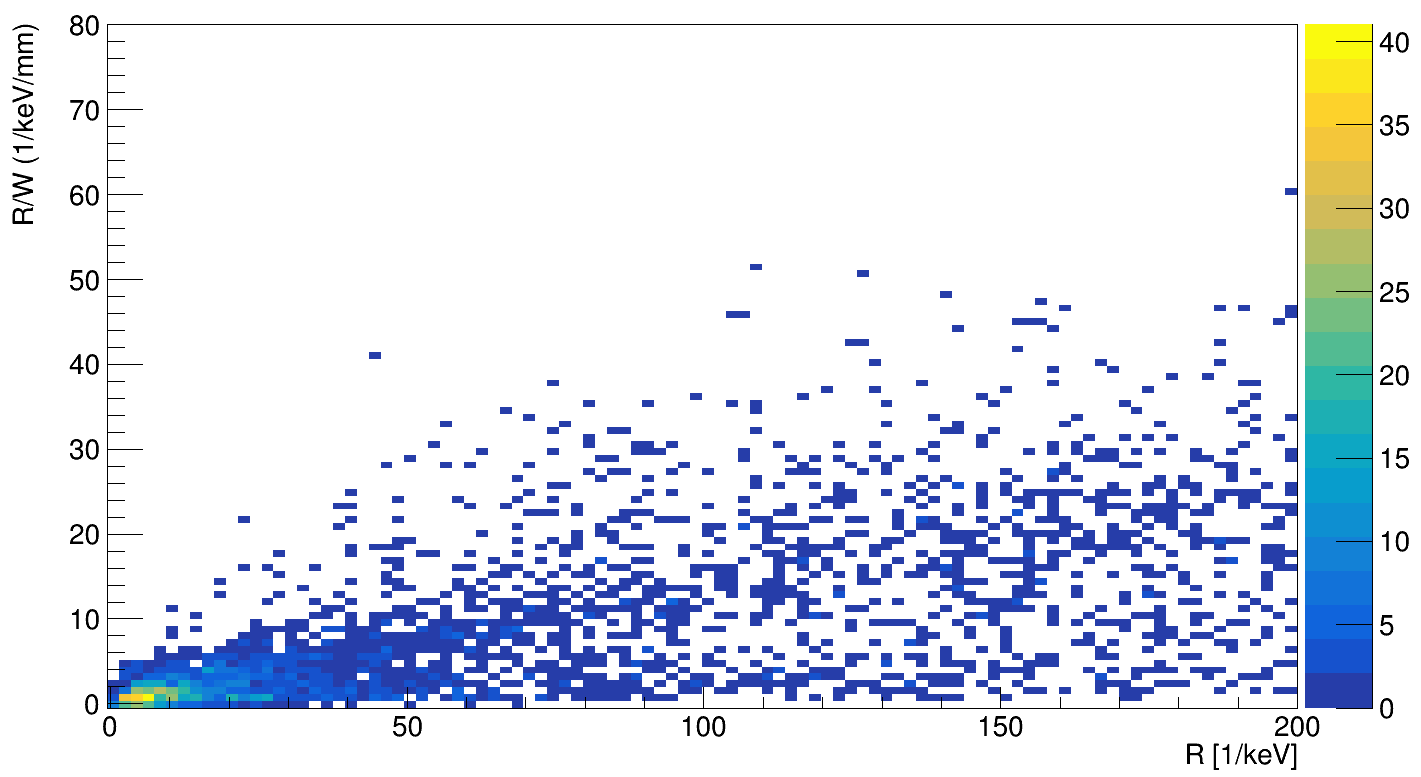} 
    \caption{Scatter plots of the ratio $R/w$ as a function of $R$ for calibration data. The figure shows a comparison between the calibration file containing only electron recoils (left) and the data file (right) before applying selection cuts. The defined cut ($R/w > 21$) is designed to exclude the electron recoil population, which dominates the lower region of the plots.}
    \label{fig:ratio_rw}
\end{figure}

To enhance the separation power, we analyze the ratio $R/w$. Nuclear recoils, being both narrow (small $w$) and dense (large $R$), occupy a distinct region in the parameter space characterized by high $R/w$ values. Conversely, the calibration data indicates that the electron recoil population is concentrated in the region where $0 < R/w < 21$ and $0 < R < 60$ as shown in figure \ref{fig:ratio_rw}.

To isolate the final nuclear recoil candidates, we apply the following selection cuts:
\begin{equation}
    R > 60 \quad \text{and} \quad R/w > 21.
\end{equation}

This cut effectively removes the residual electron background that may have survived the width selection, particularly at low energies where tracks are short and topological identification is most challenging.

The selection strategy relies on a sequence of five observables. The initial cut on the Flash-ADC derivative removes complex topologies and pile-up events. However, flash-based cuts alone are insufficient at low energies where electron tracks are short. The introduction of 3D track properties—specifically the track width $w$ and the density ratio $R/w$—allows for more robust discrimination. 


\subsection{Ionization Quenching Factor}
\begin{figure}[ht!]
    \centering
    \includegraphics[width=0.7\textwidth]{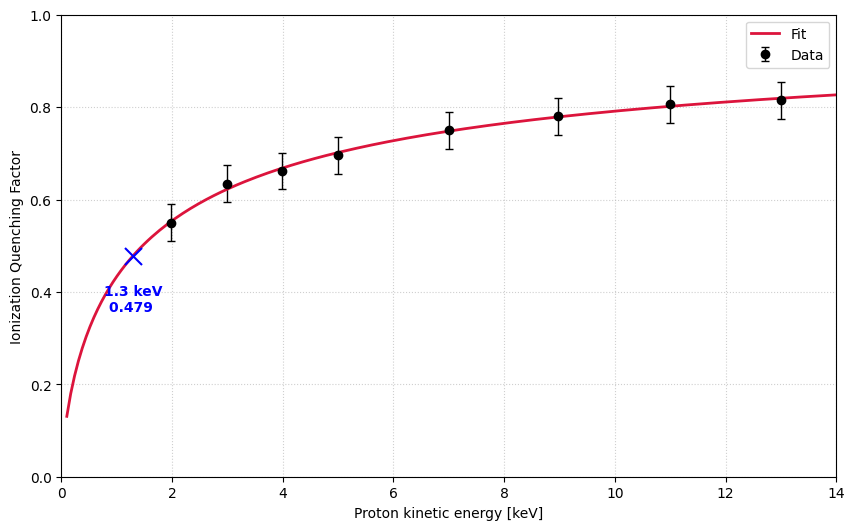}
    \caption{The measured ionization quenching factor for hydrogen recoils in CH$_4$ gas at 100 mbar \cite{balogh_2022}  using Comimac facility \cite{beaufort_2021,Muraz2016}. The red line shows a fit to the experimental data (black points).}
    \label{fig:quenching_factor}
\end{figure}

It is crucial to note that the detector measures the ionization energy, not the full kinetic energy of a recoil, due to quenching ionization effects. A significant fraction of the initial kinetic energy is dissipated via mechanisms such as non-radiative mechanisms (kinetic transfer to gas nuclei) and radiative processes (scintillation). Consequently, a 1.3~keV deuteron recoil is expected to produce an ionization signal with a measured energy significantly lower than this value, in accordance with measured quenching factors (see Figure~\ref{fig:quenching_factor}). 
As shown in Figure~\ref{fig:quenching_factor}, the data for hydrogen recoils in the CH$_4$ gas at 100 mbar indicate that a recoil with a kinetic energy of 1.3~keV would have a quenching factor of about 0.48, considering that our gas mixture is based on C$_4$H$_{10}$ gas, not CH$_4$. This means that if the recoiling particle were a proton, we would expect to measure an ionization energy of only $0.48 \times 1.3~\text{keV} \approx 0.62~\text{keV}$. 

However, the particle produced by thermal neutron capture is a deuteron. A deuteron is twice as massive as a proton, and therefore, for the same kinetic energy, it has a lower velocity. The process of quenching is strongly dependent on particle velocity, becoming more pronounced (i.e., the quenching factor decreases) for slower particles. We therefore expect the quenching factor for a 1.3~keV deuteron to be even smaller than the 0.48 value measured for a proton of the same energy. This implies that the ionization signals from thermal neutron captures should be centered at an energy below 0.62~keV.

\subsection{Final Event Selection and Identification}

\begin{figure}[ht!]
    \centering
    \includegraphics[width=0.6\textwidth]{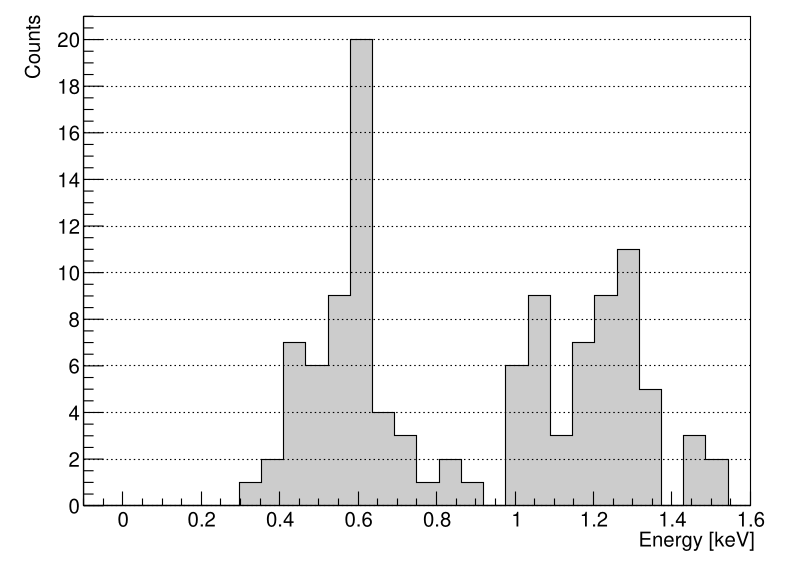}

    \caption{The energy spectrum in the sub-keV region showing the deuteron recoils from thermal neutron capture candidates close to fast neutron proton recoil events.}
    \label{fig:energy_spectrum}
\end{figure}

The application of the full set of discrimination criteria to the 
7392 minute data set allows for the isolation of a high-purity sample of nuclear recoil (NR) events. The resulting energy spectrum is presented in Figure \ref{fig:energy_spectrum}, showing the total nuclear recoil background up to 1.6~keV. To investigate the thermal neutron capture signal, we focus on the sub-keV Region of Interest (ROI), where a distinct structure is visible near 0.6~keV, sitting close to a  background of proton recoils from neutron elastic scattering.

The primary challenge lies in distinguishing the 1.3~keV deuteron recoils (signal) from the low-energy hydrogen recoil population (background). While both are nuclear recoils, their track morphology differs fundamentally due to the mass difference. According to the stopping power theory, a deuteron ($A=2$) has a higher energy loss per unit path length than hydrogen ($A=1$) of the same energy, leading to a much shorter track length ($SL$). Furthermore, ionization quenching theory indicates that the heavier the atom, the less efficient it is at producing primary electron-ion pairs. Consequently, the deuteron produces a lower number of activated pixels ($N_{pixels,3D}$) for a given kinetic energy compared to a lighter hydrogen.

\begin{figure}[ht!]
   \centering
   \includegraphics[width=0.6\textwidth]{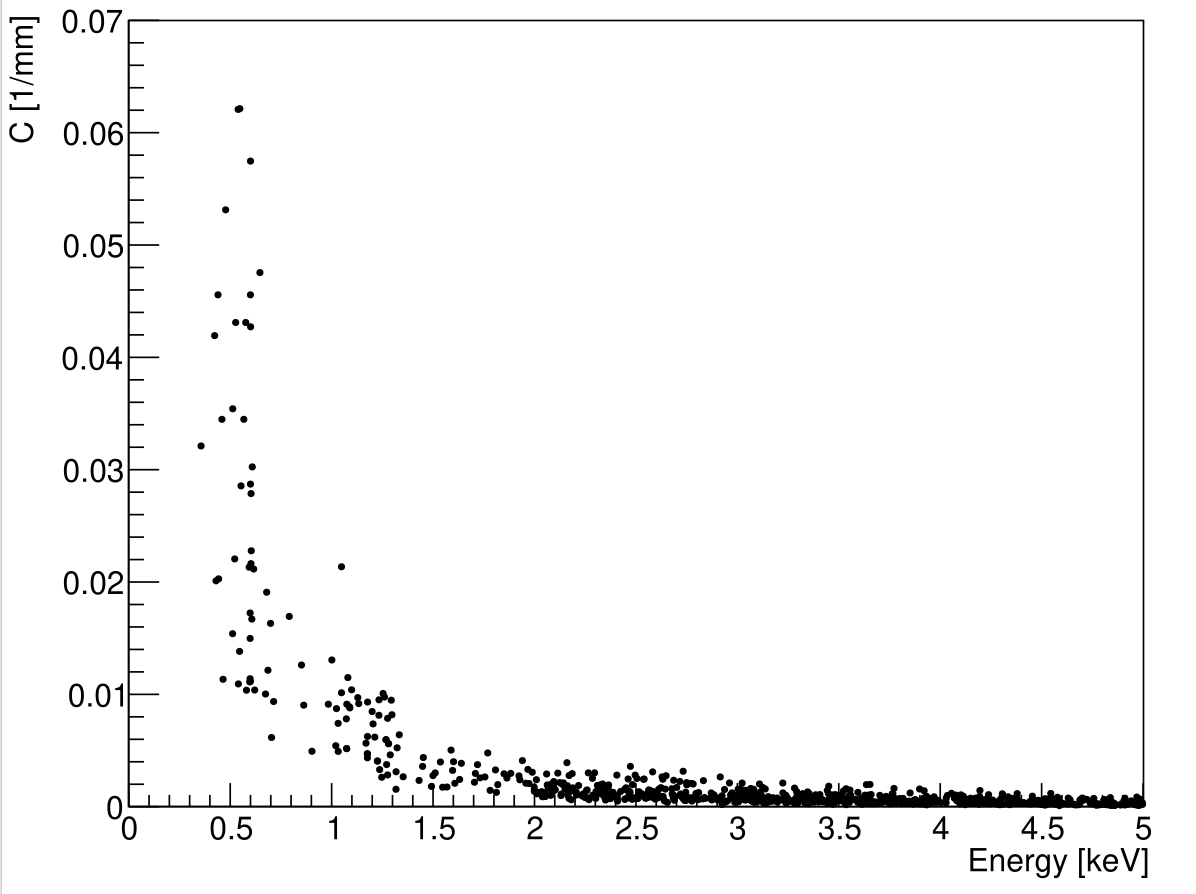} 
   \caption{Distribution of the topological observable $C$ as a function of energy. The background hydrogen population stays at low density values, whereas the deuteron signal forms a distinct high-density cluster at low energy.}
   \label{fig:bidim_density}
\end{figure}

To exploit these  physical effects, we introduced a topological observable which computes the compactness of each event, defined as $C = 1 / (N_{pixels,3D} \times SL)$. Since both the track length and the number of pixels are reduced for deuterons, this inverse product is expected to be substantially higher for the capture signal than for the hydrogen background. Figure \ref{fig:bidim_density} confirms this behavior: while background hydrogen recoils are concentrated at low values, a distinct cluster of events appears at low energy with significantly higher values, providing a clear experimental signature of the more compact deuteron tracks.

\begin{figure}[ht!]
   \centering
   \includegraphics[width=0.55\textwidth]{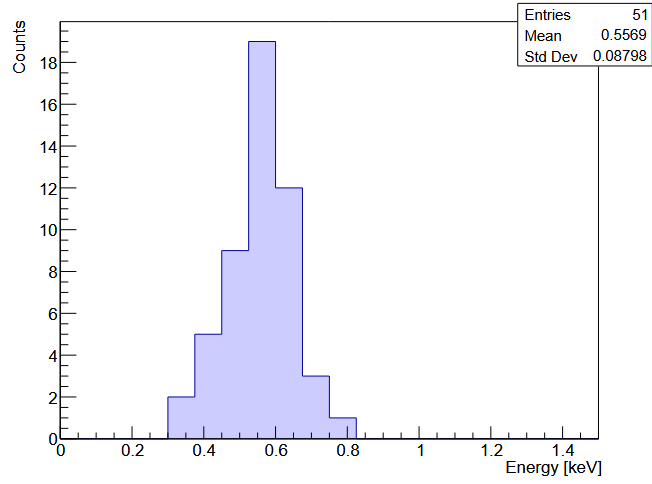} 
   \caption{Energy spectrum of the 51 events selected via the topological density cut. The peak at 0.56~keV matches the expected quenched ionization energy for 1.3~keV deuteron recoils.}
   \label{fig:final_peak}
\end{figure}

By applying a cut of $C > 0.01$~mm$^{-1}$ for energies below 0.8~keV, we identified a final sample of \textbf{51 neutron capture events}. The energy spectrum of these candidates is shown in Figure \ref{fig:final_peak}. The distribution exhibits a well-defined peak at \textbf{0.56 $\pm$ 0.09~keV}. This measured energy is consistent with the ionization yield expected from the quenching curve; because the deuteron is more massive and slower than hydrogen of the same kinetic energy, it undergoes more quenching, shifting its ionization signal to $\sim$0.56~keV.

The observed count of $51 \pm 7$ events is in agreement with our previous independent prediction. It falls within the confidence intervals of both the analytical estimate ($49 \pm 7$ events) and the PHITS Monte Carlo simulation ($61 \pm 8$ events), providing a robust cross-validation of the thermal neutron capture rate.

\subsubsection{ Proton Recoil Background from Neutron Elastic scattering}

A potential concern in the identification of the 0.56~keVee peak is the "energy degeneracy" between deuterons produced by neutron captures and proton recoils produced by neutron elastic scattering. An ionization signal of $E_{ee} \approx 0.56$~keVee can be produced through two distinct channels:
\begin{enumerate}
    \item \textbf{Thermal Capture (Signal):} A 1.3~keV kinetic energy deuteron ($E_R$) produced by thermal neutron capture ($E_n \approx 0.025$~eV). The ionization energy is $E_{ee}\approx 0.56$~keVee.
    \item \textbf{Epithermal Elastic Scattering (Background):} A proton recoil of $E_R \approx 1.2$~keV. Given a quenching factor $Q_H \approx 0.47$, the ionization energy is $E_{ee} = 1.2 \times 0.47 \approx 0.56$~keVee. Such a recoil can be produced by an epithermal neutron of energy $E_n \approx 2.4$~keV (assuming the most probable scattering angle $\theta = 45^\circ$ \cite{beaufort_2021}, where $E_R = E_n \cos^2\theta$).
\end{enumerate}

To justify that the observed peak is dominated by capture, we estimate the ratio of the interaction rate densities ($R = \phi(E_n) \times \sigma(E_n)$) for these two processes using the flux and cross-sections shown in Figure \ref{fig:flux_vs_sigma}. The rate of events is proportional to the product of the differential neutron flux $\phi(E_n)$ and the respective cross-section $\sigma(E_n)$.

The parameters extracted from the atmospheric neutron flux model and \textbf{ENDF/B-VIII.0} database  \cite{NNDC}  are summarized in Table~\ref{tab:rate_comparison}.

\begin{table}[h]
\centering
\caption{Comparison of interaction rates for signal (Capture) and potential background (Elastic).}
\label{tab:rate_comparison}
\begin{tabular}{l|c|c}
\hline
\textbf{Parameter} & \textbf{Thermal Capture (Signal)} & \textbf{Epithermal Elastic (Bkg)} \\ \hline
Neutron Energy ($E_n$) & 0.025 eV & 2.4 keV \\
Diff. Flux $\phi(E_n)$ [n/cm$^2$/s/eV] & $2.13 \times 10^{-2}$ & $1.04 \times 10^{-7}$ \\
Cross-Section $\sigma(E_n)$ [cm$^2$] & $3.34 \times 10^{-25}$ (capture) & $2.01 \times 10^{-23}$ (elastic collision) \\ \hline
\textbf{Rate Density ($\phi \cdot \sigma$)} & $\mathbf{7.13 \times 10^{-27}}$ & $\mathbf{2.08 \times 10^{-30}}$ \\ \hline
\end{tabular}
\end{table}

The probability ratio is given by:
\begin{equation}
    \mathcal{R}_{S/B} = \frac{\phi_{\text{thermal}} \cdot \sigma_{\text{cap, thermal}}}{\phi_{\text{2keV}} \cdot \sigma_{\text{ela, 2keV}}} = \frac{7.13 \times 10^{-27}}{2.08 \times 10^{-30}} \approx 3429
\end{equation}

This calculation demonstrates that for the same ionization energy of 0.56~keVee, the thermal capture process is approximately 3,400 times more probable than an elastic scattering from the epithermal neutron population (2.4 keV). Furthermore, while the capture reaction produces a mono-energetic 1.3~keV deuteron leading to a sharp peak, elastic scattering from a continuous neutron spectrum would produce a broad, flat continuum rather than the distinct Gaussian-like peak observed in Figure \ref{fig:final_peak}. Therefore, the contribution of epithermal elastic proton recoils to the identified sample of 51 events is statistically negligible.

\subsection{Directional Analysis and Isotropy}

\begin{figure}[ht!]
   \centering
   \includegraphics[width=0.4\textwidth]{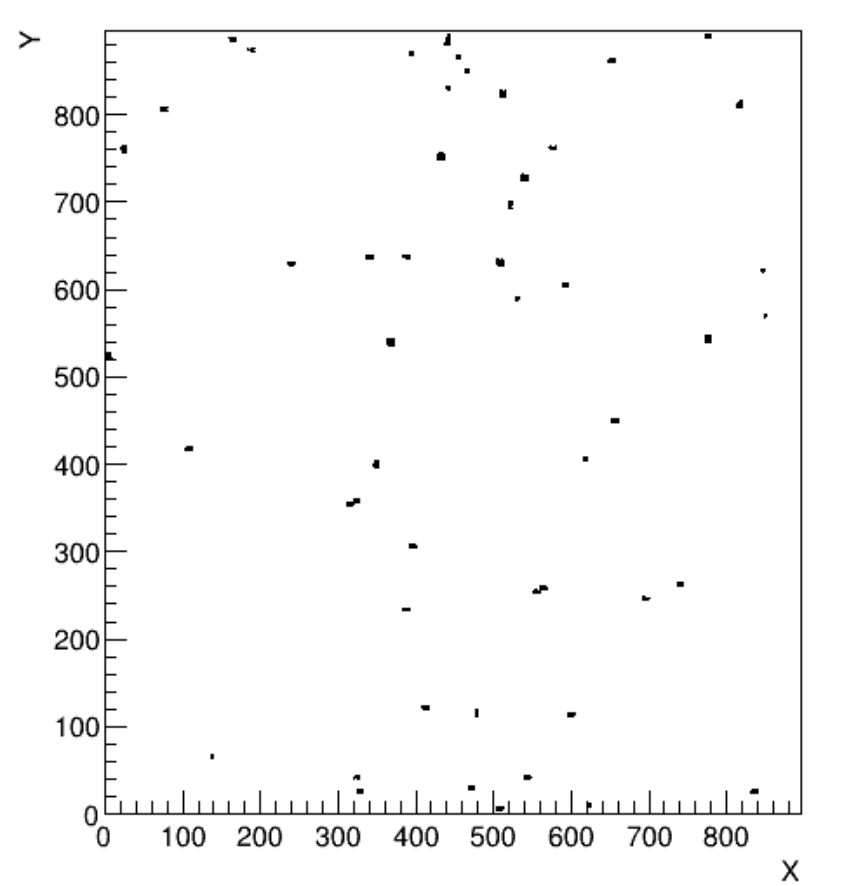} 
   \caption{Projection of the 51 identified capture events on the anode plane (XY), showing a uniform spatial distribution consistent with a diffuse background.}
   \label{fig:anode_proj}
\end{figure}

To confirm that these 51 events originate from thermal neutron capture rather than localized sources, we performed a 3D directional reconstruction. The spatial distribution of their reconstructed tracks projected on the anode plane (Figure \ref{fig:anode_proj}) shows a uniform coverage of the detector volume.

At this ionization energy scale ($E_{ee} < 0.8$~keV), the physical track length ($L \approx 750$~$\mu$m) is determined via SRIM simulation \cite{srim} and is significantly smaller than the 3D diffusion scale ($\sigma \approx 2$~mm). To accurately recover the orientations, we implemented a reconstruction method based on Maximum Likelihood Estimation (MLE). The initial linear regression results were used as first-guess parameters to initialize the fit. The track is modeled as a 3D segment of length $L$, starting at $(x_0, y_0, z_0)$ and oriented along the direction $(\theta, \phi)$. The best-fit parameters are obtained by minimizing the Negative Log-Likelihood (NLL) function:
\begin{equation}
    NLL(\theta, \phi, x_0, y_0, z_0, L) = \sum_{i=1}^{N_{voxels}} \left[ \frac{\Delta x_i^2 + \Delta y_i^2}{2\sigma_T^2} + \frac{\Delta z_i^2}{2\sigma_L^2} \right],
    \label{eq:nll}
\end{equation}
where $\Delta x_i, \Delta y_i, \Delta z_i$ represent the residuals between each voxel $i$ and the nearest point on the track segment. The diffusion widths, $\sigma_T$ and $\sigma_L$, are calculated using the \texttt{Magboltz} coefficients and the specific drift distance of the event. To ensure an unbiased measurement, the track length $L$ was treated as a free parameter during the minimization process. For an isotropic source, such as the ambient thermal neutron background, the recoil vectors are expected to be distributed uniformly across the full 4$\pi$ solid angle, with no privileged direction.



\begin{figure}[ht!]
   \centering
  \includegraphics[width=0.5\textwidth]{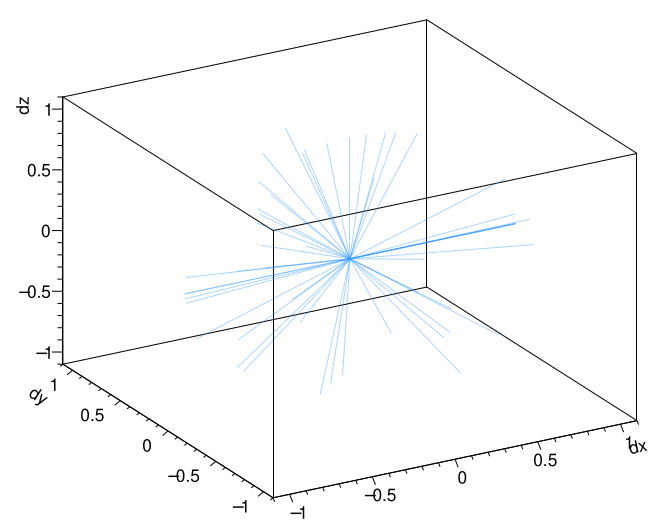}
  \caption{3D reconstruction of the 51 identified thermal neutron capture events.}
  
   \label{fig:angle}
\end{figure}

\begin{figure}[ht!]
    \centering
    \includegraphics[width=\textwidth]{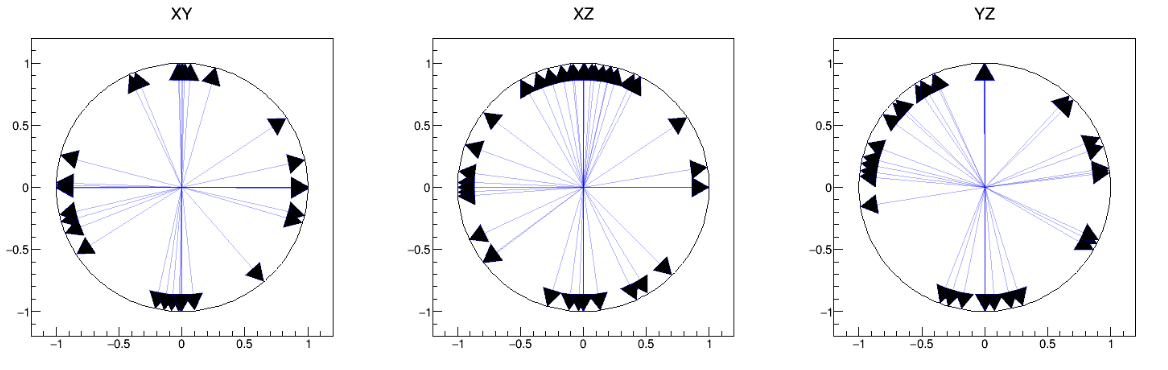}
    \caption{Two-dimensional projections of the reconstructed recoil vectors. Note: The label indicates N=51 for illustration purposes of the angular features. 
    \textbf{Left (XY Projection):} The distribution of directions in the anode plane is uniform. 
    \textbf{Center (XZ Projection) and Right (YZ Projection)}.}
    \label{fig:projections}
\end{figure}
 By minimizing the Negative Log-Likelihood function (Eq. \ref{eq:nll}) of the 51 identified events, the algorithm determines for each event the best-fit parameters $\theta$ and $\phi$ that define its 3D orientation while taking into account the diffusion widths ($\sigma_T, \sigma_L$). The resulting orientations for the 51 candidates are presented as 3D vectors in Figure \ref{fig:angle}, along with their projections onto the XY, XZ, and YZ planes in Figure \ref{fig:projections}.

The spatial distribution of these reconstructed vectors shows a uniform coverage within the detector volume, with no privileged direction observed. This isotropic signature is consistent with the expected behavior of a diffuse thermal neutron background capturing uniformly within the sensitive volume. These results demonstrate the capability of the MIMAC  detector strategy to reconstruct the 3D orientation of nuclear recoils even at the 1 keV scale, where the physical track is significantly smaller than the diffusion scale.

\section{Conclusion}

In this work, we have performed a direct measurement of thermal neutron capture on hydrogen using the MIcro-TPC MAtrix Chamber (MIMAC-35). By operating a gaseous Time Projection Chamber with a 70\% isobutane and 30\% trifluoromethane mixture at 30~mbar, we successfully isolated the low-energy nuclear recoil signature resulting from the $^{1}\text{H}(n, \gamma)^{2}\text{H}$ reaction.

Our analysis identified 51 neutron capture events over 443,519 seconds of data-taking. This result shows  statistical agreement with both our analytical estimate of $49 \pm 7$ captures and the PHITS-based Monte Carlo simulation which predicted $61 \pm 8$ captures. The identification of the recoiling deuteron peak at an ionization energy of $0.56 \pm 0.09$~keV confirms the expected quenching effects for a 1.3~keV kinetic energy recoil in this gas mixture. Furthermore, the 3D directional reconstruction of these events revealed a uniform spatial distribution and an isotropic angular distribution, consistent with an isotropic deuteron recoil production from an homogeneous thermal neutron background.

The success of this measurement relies on a robust multi-stage discrimination strategy. By using 3D track topology, specifically the track width ($w$) and the ionization density ratio ($R/w$), we demonstrated the ability to suppress dominant electron recoil (ER) background even in the sub-keV regime. This work serves as a crucial experimental validation of the MIMAC technology's performance at the 1~keV scale. The ability to precisely characterize and identify the $^{1}\text{H}(n, \gamma)^{2}\text{H}$ contribution to the background is of fundamental importance for the rare-event searches, including direct dark matter detection (WIMPs) and the observation of Coherent Elastic Neutrino-Nucleus Scattering (CEvNS), where such low-energy nuclear recoils constitute an otherwise ultimate background.

\bibliographystyle{JHEP} 
\bibliography{kha}   
\end{document}